\documentclass[10pt]{article}
\usepackage[utf8]{inputenc}

\usepackage[left=1.4in,top=1.3in,right=1.3in,bottom=1.1in,footskip = 0.333in]{geometry}

\usepackage{authblk, amsmath, amsthm, amsfonts, amssymb, amsbsy, bbm, bigstrut, graphicx, enumerate,  upref, longtable, comment, comment,xcolor,siunitx,mathrsfs}
\usepackage{float}
\usepackage{tabularx, booktabs, makecell, caption}
\usepackage{siunitx}
\usepackage{algorithm}

\usepackage{subcaption}
\usepackage[breaklinks]{hyperref}

\usepackage[numbers, sort&compress]{natbib} 
\usepackage{comment}
\usepackage[export]{adjustbox}

\newtheorem{theorem}{Theorem}

\newtheorem{prop}{Proposition}

\theoremstyle{definition}
\newtheorem{remark}{Remark}
\newtheorem{example}{Example}
\newtheorem{assn}{Assumption}

\numberwithin{theorem}{section}
\numberwithin{remark}{section}
\numberwithin{example}{section}
\numberwithin{assn}{section}
\numberwithin{defn}{section}
\numberwithin{lemma}{section}
\numberwithin{corollary}{section}
\numberwithin{prop}{section}

\newcommand{\argmin}{\operatorname{argmin}}

{\tiny {\tiny }}

\allowdisplaybreaks

\makeatletter
\def\thickhline{%
	\noalign{\ifnum0=`}\fi\hrule \@height \thickarrayrulewidth \futurelet
	\reserved@a\@xthickhline}
\def\@xthickhline{\ifx\reserved@a\thickhline
	\vskip\doublerulesep
	\vskip-\thickarrayrulewidth
	\fi
	\ifnum0=`{\fi}}
\makeatother

\newlength{\thickarrayrulewidth}
\usepackage{algorithm}

\numberwithin{equation}{section}

\renewcommand{\hat}{\widehat}
\renewcommand{\tilde}{\widetilde}

\allowdisplaybreaks

 \def \P{\mathbb{P}}

\begin{document}
\title{Late Fusion Multi-task Learning for Semiparametric Inference with Nuisance Parameters}
\date{}

\author[1]{Sohom Bhattacharya\thanks{bhattacharya.s@ufl.edu}}
\author[2]{Yongzhuo Chen \thanks{chen.yongzhuo@ufl.edu}}
\author[3]{Muxuan Liang \thanks{muxuan.liang@ufl.edu}}
\affil[1]{Department of Statistics, University of Florida}
\affil[2]{Department of Biostatistics, University of Florida}
\makeatletter
\renewcommand\AB@affilsepx{\\ \protect\Affilfont}
\renewcommand\Authsep{, }
\renewcommand\Authand{, }
\renewcommand\Authands{, }
\makeatother

\maketitle 

\begin{abstract}
In the age of large and heterogeneous datasets, the integration of information from diverse sources is essential to improve parameter estimation. Multi-task learning offers a powerful approach by enabling simultaneous learning across related tasks. In this work, we introduce a late fusion framework for multi-task learning with semiparametric models that involve infinite-dimensional nuisance parameters, focusing on applications such as heterogeneous treatment effect estimation across multiple data sources, including electronic health records from different hospitals or clinical trial data. Our framework is two-step: first, initial double machine-learning estimators are obtained through individual task learning; second, these estimators are adaptively aggregated to exploit task similarities while remaining robust to task-specific differences. In particular, the framework avoids individual level data sharing, preserving privacy. Additionally, we propose a novel multi-task learning method for nuisance parameter estimation, which further enhances parameter estimation when nuisance parameters exhibit similarity across tasks. We establish theoretical guarantees for the method, demonstrating faster convergence rates compared to individual task learning when tasks share similar parametric components. Extensive simulations and real data applications complement the theoretical findings of our work while highlight the effectiveness of our framework even in moderate sample sizes.
\end{abstract}

\section{Introduction}
\label{sec:intro}

In the modern age of machine learning, data is collected across increasingly large and heterogeneous datasets (or tasks), making the need to effectively integrate diverse sources of information crucial. Multi-task learning~\citep{zhang2018overview} has emerged as a promising approach to address this issue by leveraging the similarity across several tasks and learning them simultaneously. Since the common structure shared across tasks is unknown to a statistician, ideal multi-task learning should also be robust against possible differences across tasks while being adaptive. Our work proposes a general multi-task learning framework for semiparametric problems to improve the estimation of parameters of interest, while each task may involve similar or dissimilar parameters, as well as similar or dissimilar nuisance parameters. When tasks share similarities in these parameters, our framework facilitates faster convergence rates in estimation, offering a significant advantage in settings where task relationships are partially or fully shared.

Data fusion, i.e., the integration of information from diverse datasets, has been broadly categorized into three
approaches in literature: early, intermediate, and late fusion~\citep{baltruvsaitis2018multimodal,sidheekh2024credibility}. Early fusion combines information from
multiple sources at the input level. Intermediate fusion achieves a middle ground and attempts to jointly learn information
from both datasets in a more sophisticated manner than simply concatenating them.  Late fusion learns independently from each dataset and combines information at the end. In this work, we consider late fusion estimators based on double machine-learning, which learn independently from each dataset and then combine information to construct an improved estimation for each dataset. These algorithms differ from transfer learning and meta-learning \citep{li2022transfer, li2023transfer,tian2023transfer, cai2024transfer, zhou2024doubly}, where there is a specific target task and a statistician uses information from other data sources to improve learning for that target~\citep{hospedales2021meta,weiss2016survey}. In contrast, our work is motivated by empirical success in multi-task learning with parameter augmentation \citep{chen2011integrating,jalali2013dirty} and focuses on learning all relevant tasks rather than a specific target task.

Our late fusion multi-task learning framework offers a key advantage: privacy-preserving implementation without individual-level data sharing. This is crucial for medical applications, which motivates our work. We focus on estimating average treatment effects (ATE) or conditional average treatment effects (CATE) using data from multiple sources, such as electronic health records from different hospitals and clinical trial data across studies with identical or similar interventions. Individual hospitals and studies often have diverse patient populations but limited sample sizes. Many U.S. hospitals with fewer than 100 beds generate fewer than 80 training examples annually for rare diseases~\citep{wiens2014study}. Our framework addresses this by leveraging potential similarities in CATE across tasks to improve estimation for each hospital or study. Additionally, hospitals and studies may share similar nuisance information — comparable treatment assignment mechanisms (propensity scores) or outcome regression models — that can enhance estimation. However, privacy constraints often prohibit data sharing between hospitals or studies. This limitation necessitates our framework, which avoids direct data sharing while effectively utilizing shared information across tasks.

To address this need, our approach improves the estimation of parameters of interest in a privacy-preserving and efficient manner under semiparametric models using double machine-learning. In these models, nuisance parameters can be complex and infinite-dimensional, such as an infinite-dimensional propensity score. This research lies at the intersection of two growing bodies of literature: data integration~\citep{lenzerini2002data} and double/debiased machine learning for semiparametric models~\citep{chernozhukov2018double}. Many existing work in data integration focuses on parametric models~\citep{bastani2021predicting,duan2022heterogeneity,gu2022robust,maity2022minimax,tripuraneni2021provable}, without addressing semiparametric models involving potentially infinite-dimensional nuisance parameters. In semiparametric literature, double/debiased machine learning involving multiple nuisance parameters has been of extreme importance in the past decade~\citep{bonvini2024doubly,chernozhukov2018double,dukes2020doubly,liu2021double, feuerriegel2024causal,kennedy2023towards,nie2021quasi,wager2018estimation}, motivated by challenges in CATE estimation~\citep{bodory2022evaluating,colangelo2020double,diaz2020machine}, precision medicine~\citep{hunter2023medical,sanchez2022causal}, policy evaluation~\citep{kreif2019machine,heiler2021effect,hitsch2024heterogeneous}. However, all these works deal with the setup using data collected through one source and are not suited to handle samples from multiple sources. Our work addresses this important gap by providing a late fusion framework of adaptively aggregating estimators from individual task learning and rigorously establishing the estimation with theoretical guarantees under semiparametric models involving possibly infinite-dimensional nuisance parameters. This paper investigates the interplay between the complexity of nuisance parameters and task relatedness, focusing particularly on ATE or CATE estimation.

\textbf{{Summary of our contributions}}: Our contributions are summarized as follows.
\begin{enumerate}
    \item[(i)] We propose a general two-step late fusion framework for multi-task learning in semiparametric models with infinite-dimensional nuisance parameters. In the first step, an initial estimator is obtained from individual task learning, which can be achieved by solving estimating equations with plug-in nuisance estimates. In the second step, those initial estimators are aggregated adaptively by solving an optimization problem to leverage the similarity across tasks while being robust against outlier tasks. 
    \item[(ii)] The proposed framework does not assume any similarity among the nuisance parameters associated with the tasks. To further leverage possible similarity in the nuisance parameters (details provided by Assumption~\ref{assn:nuisance_similarity}), we propose a novel late fusion multi-task nonparametric learning for nuisance parameters. When the nuisance parameters are similar across tasks, the proposed method improves the estimation of nuisance parameters and thus improves the estimation of parameters of interest. In our implementation, we develop a pipeline that avoids direct data sharing, assuming data from each task are stored in separate local servers, and there is a central server available for late fusion.
    \item[(iii)] As our theoretical contribution, we show that the proposed late fusion multi-task learning estimators for our parameters of interest perform no worse than initial estimators obtained from individual tasks (Theorem~\ref{thm:general}). If the parameters of interest are similar across tasks, our proposed method can achieve a provably faster convergence rate than individual task learning. Further, if both the parameters of interest and nuisance parameters are similar across tasks, our proposed method can achieve an even faster convergence rate than that with nuisance parameters estimated by individual task learning (Theorem~\ref{thm:mtl_nuisance}).
    \item[(iv)] Finally, we complement our theoretical findings via extensive simulations (Section~\ref{sec:simulation}) and real-data examples (Section~\ref{sec:real_data}), where we estimate the CATE of phone consultation in mammography screening studies. Our method outperforms independent-task learning and parametric multi-task learning method common in literature.
\end{enumerate}


The rest of the paper is structured as follows. Section~\ref{sec:method} introduces our proposed framework, including multi-task learning for parameters of interest, along with examples of implementation without direct data sharing across tasks, and multi-task nonparametric learning for nuisance parameter estimation. In Section~\ref{sec:theory}, we provide a theoretical justification of the proposed methods. In Section~\ref{sec:simulation} and Section~\ref{sec:real_data}, we conduct several simulations and real data analysis to show the superiority of the proposed methods to individual task learning. Section~\ref{sec:diss} provides a discussion and future research directions. Proofs of our theoretical results and additional simulations are deferred to the Supplementary material.

\section{Methods}
\label{sec:method}

In this section, we delineate our proposed late fusion multi-task learning method for semiparametric models. We first introduce the general framework and then discuss specific examples.

\subsection{General framework}
\label{sec:general_frame}

We consider the canonical multi-task learning setting where random samples are observed from $K$ different
populations. The $i$-th observation from the $k$-th population satisfies
\begin{align}\label{eq:model}
    Z_{k,i} \sim h(z, \theta_{k}, \eta_k), \quad i \in \{1,\ldots, n\}, \quad k \in \{1,\ldots, K\},
\end{align}
where $(\theta_{k}, \eta_k)$ are task-specific parameters, and $h(z, \cdot, \cdot)$ is a probability density function. Throughout the paper, $\theta_k$ denotes the parameters of interest, and $\eta_k$ denotes the possibly infinite dimensional nuisance parameters. We assume throughout that an equal number of samples $n$ is observed from each population --- this is for notational convenience and can be easily extended for unequal sample sizes. The objective of this paper is to learn $\theta_k$'s simultaneously rather than focusing on a specific task. 

The general structure of the late-fusion estimator is outlined in Algorithm~\ref{algo:general}. In this procedure, the first-step estimators $\widetilde{\theta}_k$ are obtained via individual task learning. The function $\psi_1(\cdot, \cdot)$ denotes a loss associated with the chosen estimating equations (defined below), while $\psi_2(\cdot, \cdot)$ is a distance metric that captures task similarity. We will specify our choices for $\tilde \theta_k$, $\psi_1$, $\psi_2$ in semiparametric problems. The weights $w_k$ are necessary to account for the possible unequal sample sizes across different tasks.

\begin{algorithm}
\small
	\caption{Late fusion estimator}
\textbf{Input:} $Z_{k,i}$, weights $w_k$, $i \in \{1,\ldots, n\}, k \in \{1,\ldots, K\}$, tuning parameter $\lambda$, distance $\psi_1,\psi_2$.

\textbf{Output:} $\widehat{\theta}_1, \ldots, \widehat{\theta}_k$, two-step estimators of $\theta_1,\ldots, \theta_k$.

i. Obtain initial estimators $\widetilde{\theta}_k$ based on $k$-th sample, $k \in \{1,\ldots, K\}$.

ii. Aggregate the initial estimators by solving the following optimization problem:
\begin{equation}\label{eq:armul_gen}
    (\widehat{\theta}_0,\ldots, \widehat{\theta}_K)= \argmin_{u_{0},\ldots, u_K} \sum_{k=1}^K w_k \Big\{\psi_1(u_k, \widetilde{\theta}_k)+\lambda \psi_2(u_k,u_0)\Big\}
\end{equation}
\quad The final estimators are $(\widehat{\theta}_1,\ldots, \widehat{\theta}_K)$.\label{algo:general}
    \end{algorithm}

Throughout the paper, we make the standard assumption following~\citet{chernozhukov2018double} that $\theta_k$'s can be identified through a moment function $E_k\left[m(Z, \theta,\eta_k)\right]=0$, i.e., $\theta_k$ is the unique solution of $E_k\left[m(Z, \theta,\eta_k)\right]=0$, where $E_k[\cdot]$ is the population mean corresponding to the $k$-th population. A direct approach to estimate $\theta_k$'s is the individual task learning. Individual task learning solves the empirical version of the moment function with a plug-in estimator for $\eta_k$ using only the samples from the $k$-th population, i.e., 
\begin{equation}\label{eq:define_score}
    \widehat{E}_k\left[m(Z, \theta,\widetilde{\eta}_k)\right]=0,
\end{equation}
where $\widetilde{\eta}_k\equiv \widetilde{\eta}_k(Z,\theta)$ is a plug-in estimator for $\eta_k$, and $\widehat{E}_k[\cdot]$ is referred to as the empirical average of samples from the $k$-th population. However, individual task learning overlooks the similarity among tasks. In this work, we propose late fusion estimators, which exploit the similarity among $\theta_k$'s and ${\eta}_k$'s to improve our estimation. Before describing our aggregation procedure, we provide some examples of estimating equations and initial estimators we will consider.

\begin{example}\label{example:1}
Let $Z_{k,i}=\left(X_{k,i}, T_{k,i}, Y_{k,i}\right)$, where $Y_{k,i}$ is an outcome of interest, $X_{k,i} \in \mathbb R^p$ be the covariate vector, $T_{k,i}$ is a single covariate whose effect is of interest. Consider the partial linear regression model across $K$ different tasks:
\begin{align*}
    &Y_{k,i}= T_{k,i} \theta_k + f_k(X_{k,i})+ \varepsilon_{k,i}, \quad T_{k,i}= g_{k}(X_{k,i})+ \eta_{k,i}, \quad i \in \{1,\ldots, n\}, \quad k \in \{1,\ldots, K\}.
\end{align*}
Suppose $\varepsilon_{k,i}$'s are i.i.d. with mean $0$, variance $\sigma^2_{\varepsilon}$, $\eta_{k,i}$'s are i.i.d. with mean $0$, variance $\sigma^2_{\eta}$. For individual task learning, efficient estimations of $\theta_k$'s have been well-studied in \citet{chernozhukov2018double,dukes2021inference}. Here, we use the estimating equation:
\begin{eqnarray}\label{eq:plm_equation}
    m(Z_{k,i}, \theta,{\eta}_k)=\left\{Y_{k,i}-\mu_k(X_{k,i})-\left(T_{k,i}-g_k(X_{k,i})\right)\theta\right\}\left(T_{k,i}-g_k(X_{k,i})\right),
\end{eqnarray}
where $\eta_k=(\mu_k(\cdot),g_k(\cdot))$ and $\mu_k(X_{k,i})=E[Y_{k,i}\mid X_{k,i}]$. Individual task learning first obtains $\widetilde{\eta}_k$ and then solves $\widehat{E}_k\left[m(Z, \theta,\widetilde{\eta}_k)\right]=0$. The cross-fitting procedure is adopted to accommodate $\widetilde{\eta}_k$ estimated using machine learning or nonparametric methods. Our framework leverages the similarity among $\theta_k$'s and $\eta_k$'s to improve the estimation of each $\theta_k$.
\end{example}

\begin{example}\label{example:2}
Let $Z_{k,i}=\left(X_{k,i}, Y_{k,i}\right)$, where $Y_{k,i}$ is a outcome of interest, and $X_{k,i}$ is a $p$-dimensional covariate vector. Consider a single-index regression model across $K$ different tasks:
\begin{align*}
    &Y_{k,i}= f_k(X_{k,i}^\top \theta_k)+ \varepsilon_{k,i}, \quad i \in \{1,\ldots, n\}, \quad k \in \{1,\ldots, K\}.
\end{align*}
Suppose $\varepsilon_{k,i}$'s are i.i.d. with mean $0$ and variance $\sigma^2_{\varepsilon}$. Note that we allow the link functions $f_k$'s to be potentially different across tasks. To ensure the identification of $\theta_k$'s, we assume the first coordinates of $\theta_k$'s equal to $1$. For individual task learning, efficient estimations of $\theta_k$'s adopt an estimating equation \citet{ma2012semiparametric, ma2013efficient} as
\begin{eqnarray*}
    m(Z_{k,i}, \theta,{\eta}_k)=\left\{Y_{k,i}-f_k\left(X_{k,i}^\top\theta\right)\right\}f_k^{'}\left(X_{k,i}^\top\theta\right)\left(\widetilde{X}_{k,i}-g_k\left(X_{k,i}^\top\theta\right)\right),
\end{eqnarray*}
where $\widetilde{X}_{k,i}$ equals ${X}_{k,i}$ after removing the first coordinate, $\eta_k=(g_k(\cdot), f_k(\cdot))$, and $g_k\left(X_{k,i}^\top\theta\right)=E\left[X_{k,i}\mid X_{i,k}^\top\theta\right]$. 
\end{example}

\begin{example}\label{example:3}
Let $Z_{k,i}=\left(X_{k,i}, T_{k,i}, Y_{k,i}\right)$, where $Y_{k,i}$ is the outcome of interest, $X_{k,i}$ is a $p$-dimensional covariate vector, $T_{k,i}\in\{-1, 1\}$ is a binary treatment whose effect we want to estimate. Assume causal assumptions described in the online Supplementary Material. Consider the single-index model across $K$ different tasks:
\begin{align*}
    &Y_{k,i}= \frac{1}{2}T_{k,i} f_k\left(X_{k,i}^\top\theta_k\right) + \mu_k(X_{k,i})+ \varepsilon_{k,i}, \quad i \in \{1,\ldots, n\}, k \in \{1,\ldots, K\}.
\end{align*}
In this model, $f_k\left(X_{i,k}^\top\theta_k\right)$ represents the CATE, which is of our major interest, and $\mu_k(X_{k,i})$ represents the main effect which contributes to the outcome no matter which treatment option is assigned.
For individual task learning, estimations of $\theta_k$'s can be achieved through methods in dimension reduction, e.g., \citet{ma2012semiparametric, ma2013efficient, liang2022semiparametric}, which are motivated by estimating equations as
\begin{eqnarray*}
    m(Z_{k,i}, \theta,{\eta}_k)=\frac{T_{k,i}}{\pi_{k}(T_{k,i}, X_{k,i})}\left\{Y_{k,i}-\mu_k(X_{k,i})-\frac{1}{2}T_{k,i}f_k\left(X_{k,i}^\top\theta\right)\right\}f_k^{'}\left(X_{k,i}^\top\theta\right)\left(\widetilde{X}_{k,i}-g_k\left(X_{k,i}^\top\theta\right)\right),
\end{eqnarray*}
where $\eta_k=(\mu_k(\cdot), \pi_k(\cdot, \cdot),g_k(\cdot), f_k(\cdot))$, the mean function $$\mu_k(X_{k,i})=\left(E[Y_{k,i}\mid T_{k,i}=1, X_{k,i}]+E[Y_{k,i}\mid T_{k,i}=-1, X_{k,i}]\right)/2,$$ and the propensity scores $\pi_{k}(T_{k,i}, X_{k,i})=P_k(T_{k,i}\mid X_{k,i})$. 
\end{example}

\begin{remark}[Neyman near-orthogonality]
In this work, we focus on $m(., \theta, \eta_k)$ that satisfy the Neyman near-orthogonality property, as detailed in Assumption~\ref{assn:true_theta}. While our framework can be applied to any consistent estimating equation without this requirement, Neyman near-orthogonal estimating equations are particularly advantageous because they minimize the impact of nuisance parameter estimation errors on estimating parameters of interest. For this reason, we restrict our discussion to estimating equations that meet the conditions outlined in Assumption~\ref{assn:true_theta}.
\end{remark}

\subsection{Late fusion multi-task learning for parametric components} Our framework leverages the similarity among $\theta_k$'s and $\eta_k$'s to improve the estimation of each $\theta_k$. We first introduce a late-fusion framework that can leverage the similarity among $\theta_k$'s; and then extend it to leverage the similarity among nuisance components $\eta_k$'s (see Section~\ref{sec:mtl_nonpar}). Our key idea is to aggregate initial estimators from individual task learning with a penalty that adaptively fuses similar estimators. Denote the initial estimators from individual task learning as $\widetilde{\theta}_k$'s. We consider the following optimization problem:
\begin{eqnarray}\label{eq:multi-task_parametric}
    \min_{u_{0},\ldots, u_K} \sum_{k=1}^K w_k \Big\{\rho_k(u_k, \widetilde{\theta}_k;\widetilde{\eta}_k)+\lambda \left\|u_k-u_0\right\|_2\Big\},
\end{eqnarray}
where $\rho_k$ is a suitably chosen loss function that depends on the choice of $m(Z_{k,i},\theta,\eta_k)$'s, and $\left\|\cdot\right\|_2$ denotes the $\ell_2$ norm between vectors. Denote its minimizer as $\widehat{\theta}_k$'s. The penalty term encourages similarity between the resulting estimators, i.e., $u_k$'s, while the tuning parameter $\lambda$ guards against possible outliers. Note that, if $\lambda=0$, the resulting estimators correspond to individual task learning, and if $\lambda= \infty$, the penalty term imposes $\widehat{\theta}_1 = \ldots = \widehat{\theta}_k$, similar to methods in distributed learning~\citep{duan2022heterogeneity,jordan2019communication}. The weights $w_k$'s are used to deal with unequal sample sizes across tasks and can be chosen as $1$ if the sample sizes are similar across tasks.

\textbf{Choice of loss function $\rho_k$:} The loss function $\rho_k$ in the optimization problem~\eqref{eq:multi-task_parametric} is constructed based on the estimating equation $E[m(Z_{k,i}, \theta,{\eta}_k)]=0$. Consider the following one-step Taylor's series approximation of $\widehat{E}_k\left[m(Z_{k,i}, \theta,\widetilde{\eta}_k)\right]$ around the initial estimator $\widetilde{\theta}_k$ obtained from individual tasks,
\begin{eqnarray*}
    \widehat{E}_k\left[m(Z_{k,i}, \theta,\widetilde{\eta}_k)\right]\approx \widehat{E}_k\left[m(Z_{k,i}, \widetilde{\theta}_k,\widetilde{\eta}_k)\right]+\left\{\widehat{E}_k\left[\nabla m(Z_{k,i}, \widetilde{\theta}_k,\widetilde{\eta}_k)\right]\right\}^\top\left(\theta-\widetilde{\theta}_k\right),
\end{eqnarray*}
where $\widehat{E}_k[\cdot]$ is the empirical average over the samples in the $k$-th task, and the $j$-th row of $\widehat{E}_k\left[\nabla m(Z_{k,i}, \widetilde{\theta}_k,\widetilde{\eta}_k)\right]$ is the gradient of the $j$-th element of $\widehat{E}_k\left[m(Z_{k,i}, \theta,\widetilde{\eta}_k)\right]$ w.r.t. $\theta$ at $\theta=\widetilde{\theta}_k$. Motivated by this approximation, we consider the following loss function
\begin{eqnarray}\nonumber
     \rho_k(u_k, \tilde \theta_k; \widetilde{\eta}_k)&=&\left\{\widehat{E}_k\left[m(Z_{k,i}, \widetilde{\theta}_k,\widetilde{\eta}_k)\right]\right\}^\top u_k+\frac{1}{2}\left(u_k-\widetilde{\theta}_k\right)\left\{\widehat{E}_k\left[\nabla m(Z_{k,i}, \widetilde{\theta}_k,\widetilde{\eta}_k)\right]\right\}^\top\left(u_k-\widetilde{\theta}_k\right)\\
     &\equiv&G_k^\top u_k+\frac{1}{2}u_k W_k u_k, \label{eq:define_rho}
\end{eqnarray}
where we introduce the notations
\begin{eqnarray*}
    G_k&=&\left\{\left\{\widehat{E}_k\left[m(Z_{k,i}, \widetilde{\theta}_k,\widetilde{\eta}_k)\right]\right\}^\top-\widetilde{\theta}_k^\top\widehat{E}_k\left[\nabla m(Z_{k,i}, \widetilde{\theta}_k,\widetilde{\eta}_k)\right] \right\}^\top,\\
    W_k&=&\left\{\widehat{E}_k\left[\nabla m(Z_{k,i}, \widetilde{\theta}_k,\widetilde{\eta}_k)\right]\right\}^\top.
\end{eqnarray*}
Note that $\rho_k$ defined by~\eqref{eq:define_rho} is a quadratic loss function. Under this choice of the loss function, the gradient of the loss function w.r.t. $u_k$ is the approximation of $\widehat{E}_k\left[m(Z_{k,i}, \theta,\widetilde{\eta}_k)\right]$ with $\theta=u_k$. This choice has two major benefits. First, it is strictly convex when $\widehat{E}_k\left[\nabla m(Z_{k,i}, \widetilde{\theta}_k,\widetilde{\eta}_k)\right]$ is positive definite; thus easy to optimize. Second, the loss function (as well as its gradient and hessian) can be calculated for each task separately, and thus, no individual-level data sharing across different tasks is needed, which enables an implementation that preserves privacy.

\begin{remark}[Implementation and choice of tuning parameter]\label{remark:tuning}
    Assuming the data collected from each task are stored on different (local) servers and there is a central server available, the proposed privacy-preserving implementation is as follows: we first split the samples from each task into $J$ folds. Using the samples excluded in $j$-th fold, we implement the individual task learning to obtain $\widetilde{\theta}_k^{(j)}$'s and $\widetilde{\eta}_k^{(j)}$. Then, using the samples in the $j$-th fold, we construct $\rho_k(u_k, \tilde \theta_k^{(j)}; \widetilde{\eta}_k^{(j)})$ for each task. We aggregate the gradient and hessian information of $\rho_k(u_k, \tilde \theta_k^{(j)}; \widetilde{\eta}_k^{(j)})$ calculated on each fold and transmit them to the central server. On the central server, we solve the optimization~\eqref{eq:multi-task_parametric} given a pre-specified $\lambda$ and obtain $\widehat{\theta}_k$'s. To tune $\lambda$, we implement a $J$-fold cross-validation procedure for the selection of $\lambda$ (for numerical examples, we choose five-fold cross-validation). Specifically, we can leave one of the $J$ folds out and use it for validation. For a given $\lambda$, we can obtain the estimates $\widehat{\theta}_k$'s on the central server and transmit it to the local server. Then we calculate the loss function $\rho_k(u_k, \tilde \theta_k^{(j)}; \widetilde{\eta}_k^{(j)})$ using the samples in the leave-one-out fold on the $k$-th local server. Finally, we can choose the $\lambda$ that minimizes the averaged loss among all tasks.
\end{remark}

\subsection{Example: partial linear model}
\label{sec:plm_example}

In this section, we use partial linear model as an example to illustrate the framework we proposed in the previous section. We borrow the notation and set-up described in Example~\ref{example:1}. We leave the detailed description of Examples~\ref{example:2} and~\ref{example:3} in the Supplementary Material. 

To implement our method, we first split the samples from each task into $J$ folds, denoted as $I_1, \cdots, I_J$. For the $j$-th fold, individual task learning is performed using the samples excluded from that fold. Within this process, a cross-fitting approach can also be applied: the samples excluded from the $j$-th fold are further divided into two folds, $M_1^j$ and $M_2^j$. We estimate $\mu_k(\cdot)$ and $g_k(\cdot)$ by regressing $Y_{k,i}$ and $T_{k,i}$ on $X_{k,i}$ using the data from these folds. Various nonparametric and machine learning methods can be used in these estimations; in our simulations and real data application, we employ kernel regression --
\begin{eqnarray}\label{eq:kernel_reg_plm}
    \widetilde{\mu}^{(j,m)}_k(x)=\frac{\sum_{i\not\in M_m^j\cup I_j} Y_{k,i}H_{\hbar}(X_{k,i}-x)}{\sum_{i\not\in M_m^j\cup I_j} H_{\hbar}(X_{k,i}-x)},\quad \widetilde{g}^{(j,m)}_k(x)=\frac{\sum_{i\not\in M_m^j\cup I_j} T_{k,i}H_{\hbar}(X_{k,i}-x)}{\sum_{i\not\in M_m^j\cup I_j} H_{\hbar}(X_{k,i}-x)},
\end{eqnarray}
where $H_{\hbar}(x)={\hbar}^{-p}H(x/{\hbar})$, $H(\cdot)$ is a kernel function, and ${\hbar}$ is the bandwidth. Common kernel functions include Gaussian kernel functions, Epanechnikov Kernel functions, etc. In our implementation, we use Gaussian kernel functions, and the bandwidth is tuned using task-specific cross-validation. After obtaining $\widetilde{\mu}^{(j,m)}_k(\cdot)$ and $\widetilde{g}^{(j,m)}_k(\cdot)$, we adopt the estimating equation~\eqref{eq:plm_equation} and solve it using the samples in the other fold, i.e., our initial estimator is the solution of
\begin{eqnarray*}
    \frac{1}{|M_1^j|}\sum_{i\in M_1^j} m(Z_{k,i}, \theta, \widetilde{\eta}_k^{(j,2)})+\frac{1}{|M_2^j|}\sum_{i\in M_2^j} m(Z_{k,i}, \theta, \widetilde{\eta}_k^{(j,1)})=0,
\end{eqnarray*}
where $\widetilde{\eta}_{k}^{j,m}=(\widetilde{\mu}^{(j,m)}_k(\cdot), \widetilde{g}^{(j,m)}_k(\cdot))$. We denote its solution as $\widetilde{\theta}_k^{(j)}$. The individual task learning can be implemented on local servers and does not need data sharing between local servers.
\begin{remark}[Corss-fitting]
    Our method does not require initial estimators to be computed using cross-fitting. Theorem~\ref{thm:general} establishes how the error rates of the late-fusion estimators depend on the estimation errors of the initial estimators. When cross-fitting is used to construct the initial estimators, Theorem~\ref{thm:general} directly implies the results in Theorem~\ref{thm:plm}.
\end{remark}

After the individual task learning, we calculate the first-and second-order derivatives of $\rho_k(u_k, \widetilde{\theta}_k, \widetilde{\eta}_k)$ w.r.t. $u_k$, i.e.,
\begin{eqnarray*}
    G_k&=&\frac{1}{n}\sum_{j=1}^J \left[\sum_{i\in I_j} m(Z_{k,i},\widetilde{\theta}_k^{(j)}, \widetilde{\eta}_k^{(j)})-\left\{\sum_{i\in I_j} \nabla m(Z_{k,i},\widetilde{\theta}_k^{(j)}, \widetilde{\eta}_k^{(j)})\right\}^\top \widetilde{\theta}_k^{(j)}\right],\\
    W_k&=& \frac{1}{n}\sum_{j=1}^J \left\{\sum_{i\in I_j} \nabla m(Z_{k,i},\widetilde{\theta}_k^{(j)}, \widetilde{\eta}_k^{(j)})\right\}^\top,
\end{eqnarray*}
where $\widetilde{\eta}_k^{(j)}=\left(\widetilde{\eta}_k^{(j,1)}+\widetilde{\eta}_k^{(j,2)}\right)/2$.
Then, we transmit $G_k$'s and $W_k$'s to the central server and solve~\eqref{eq:multi-task_parametric} by solving the following optimization problem:
\begin{eqnarray}\label{eq:plm_opt}
    (\widehat{\theta}_0, \ldots, \widehat{\theta}_k)= \argmin_{u_{0},\ldots, u_K} \sum_{k=1}^K w_k \left\{G_k^\top u_k+\frac{1}{2}u_k^\top W_k u_k+\lambda \left\|u_k-u_0\right\|_2\right\}.
\end{eqnarray}
The output $(\widehat{\theta}_1,\ldots,\widehat{\theta}_K)$ is the final late-fusion multi-task estimator for $\theta_1,\ldots, \theta_K$. As mentioned above, to select appropriate $\lambda$, we leave one $j$-fold out for validation and select $\lambda$ that minimizes the averaged loss over tasks.

\subsection{Improved estimation through late fusion for nuisance parameters}
\label{sec:mtl_nonpar}

In Section~\ref{sec:general_frame}, we introduce our general framework on late fusion multi-task learning for the parameters of interest in semiparametric models. Our procedure is built on moment functions that identify the parameter of interest, $m(Z,\theta_k,\eta_k)$. In the estimation of $\theta_k$'s, as shown in Section~\ref{sec:general_frame} and~\ref{sec:plm_example}, we need to construct estimators for nuisance parameters $\eta_k$'s and then estimate $\theta_k$'s with these estimators for $\eta_k$'s plugged-in. Naturally, the estimation errors of $\widetilde{\eta}_k$'s affect the estimation of $\theta_k$'s, especially when the sample size of each task is relatively small. This precise impact is theoretically characterized in Theorem~\ref{thm:general} through the parameters $r_n$ and $r_n'$ (defined in Assumption \ref{assn:score}); see detailed discussion in Section~\ref{sec:theory}. Therefore, it is important to develop improved estimation methods for nuisance parameters that leverage possible similarities between $\eta_k$'s.

In this section, we address this question by developing late fusion multi-task learning for nuisance parameter estimation, which further improves the estimation error of $\theta_k$'s. To this end, we consider using kernel regressions to estimate $\eta_k$. To illustrate our idea, we focus on estimating $\mu_k$ in Example~\ref{example:1}. Same idea and procedure can be adopted to estimate $g_k$ in Example~\ref{example:1}, $\mu_k$ and $\pi_k$ in Example~\ref{example:3}. Suppose we are interested in estimating $\mu_k(x)$ for any $x\in \mathcal{X}$. The kernel regressions~\eqref{eq:kernel_reg_plm} can be written as the minimizer of
\begin{eqnarray*}
    \frac{\sum_{i\not\in M_m^j\cup I_j} \left(Y_{k,i}-u_k\right)^2 H_{\hbar}(X_{k,i}-x)}{\sum_{i\not\in M_m^j\cup I_j} H_{\hbar}(X_{k,i}-x)},
\end{eqnarray*}
where recall that $H_{\hbar}(x)={\hbar}^{-p}H(x/{\hbar})$. Based on this observation, we consider the following optimization problems:
\begin{eqnarray}\label{eq:multi-task_nuisance}
    \min_{u_{0},\ldots, u_K} \sum_{k=1}^K w_k \left\{\frac{\sum_{i\not\in M_m^j\cup I_j} \left(Y_{k,i}-u_k\right)^2 H_{\hbar}(X_{k,i}-x)}{\sum_{i\not\in M_m^j\cup I_j} H_{\hbar}(X_{k,i}-x)}+\widetilde{\lambda} \left|u_k-u_0\right|\right\},
\end{eqnarray}
where the bandwidth $\hbar$ and tuning parameter $\widetilde{\lambda}$ can be tuned by cross-validation. Similar to~\eqref{eq:multi-task_parametric}, the penalty term encourages similarity between $u_k$'s, while the tuning parameter $\widetilde{\lambda}$ guards against possible outliers.

However, unlike~\eqref{eq:multi-task_parametric} where only finite-dimensional parameters are of interest, our target in~\eqref{eq:multi-task_nuisance} is an infinite-dimensional nuisance parameter $\mu_k(\cdot)$, i.e., our target is the value of $\mu_k(\cdot)$ at all possible values of $X$ rather than a fixed value. In addition, we want to avoid data sharing among local servers. To achieve both, we leverage the optimization~\eqref{eq:multi-task_nuisance} and propose an algorithm without directly sharing local data. First, we construct a $\hbar^2$-covering of $\mathcal{X}$, denoted as $\mathcal{T}=\left\{t_1, \cdots, t_{|\mathcal{T}|}\right\}$. Given a bandwidth $\hbar$ and tuning parameter $\widetilde{\lambda}$, we consider
\begin{eqnarray*}
    \min_{u_{0}^t,\ldots, u_K^t, \forall t\in \mathcal{T}} \sum_{t\in \mathcal{T}}\sum_{k=1}^K w_k \left\{\frac{\sum_{i\not\in M_m^j\cup I_j} \left(Y_{k,i}-u_k^t\right)^2 H_{\hbar}(X_{k,i}-t)}{\sum_{i\not\in M_m^j\cup I_j} H_{\hbar}(X_{k,i}-t)}\right\}+\widetilde{\lambda}\sum_{k=1}^K w_k \left\|u_k-u_0\right\|_2,
\end{eqnarray*}
where $u_k=\left(u_k^{t_1}, \cdots, u_k^{t_{|\mathcal{T}|}}\right)^\top$ and $u_0=\left(u_0^{t_1}, \cdots, u_0^{t_{|\mathcal{T}|}}\right)^\top$.

To solve this optimization without individual-level data sharing, we calculate the first-and second-order derivatives of the loss function w.r.t. $u_k$ on each local server, i.e., we compute
\begin{eqnarray*}
    \widetilde{G}_k&\equiv& \left(-\frac{\sum_{i\not\in M_m^j\cup I_j} Y_{k,i} H_{\hbar}(X_{k,i}-t_1)}{\sum_{i\not\in M_m^j\cup I_j} H_{\hbar}(X_{k,i}-t_1)}, \cdots, -\frac{\sum_{i\not\in M_m^j\cup I_j} Y_{k,i} H_{\hbar}(X_{k,i}-t_{|\mathcal{T}|})}{\sum_{i\not\in M_m^j\cup I_j} H_{\hbar}(X_{k,i}-t_{|\mathcal{T}|})}\right)^\top,\\
    \widetilde{W}_k&\equiv& I_{|\mathcal{T}|\times |\mathcal{T}|},
\end{eqnarray*}
where $I_{|\mathcal{T}|\times |\mathcal{T}|}$ is the identity matrix with dimension $|\mathcal{T}|\times |\mathcal{T}|$. Then these quantities $\widetilde{G}_k$ and $\widetilde{W}_k$ are transmitted to the central server and the following optimization problem is solved:
\begin{eqnarray*}
    \min_{u_{0},\ldots, u_K} \sum_{k=1}^K w_k \left\{\widetilde{G}_k^\top u_k+\frac{1}{2}u_k^\top \widetilde{W}_k u_k+\widetilde{\lambda} \left\|u_k-u_0\right\|_2\right\}.
\end{eqnarray*}
Denote the minimizer as $\widehat{\mu}_k^{(j,m)}=\left(\widehat{\mu}_k^{(j,m)}(t_1), \cdots, \widehat{\mu}_k^{(j,m)}(t_{|\mathcal{T}|})\right)^\top$'s. To estimate $\mu_k(X_{k,i})$, we transmit $\widehat{\mu}_k^{(j,m)}$ to each local server. To predict the value of $\mu_k$ at each $X_{k,i}$, we find $t\in \mathcal{T}$ nearest to $X_{k,i}$ and then use $\widehat{\mu}_k^{(j,m)}(t)$ as the estimate for $\mu_k(X_{k,i})$. After obtaining $\widehat{\mu}_k^{(j,m)}$, we obtain $\widehat{g}_k^{(j,m)}$; then we can construct $\widehat{\eta}_{k}^{j,m}=(\widehat{\mu}^{(j,m)}_k(\cdot), \widehat{g}^{(j,m)}_k(\cdot))$. Finally, we implement the proposed method in Section~\ref{sec:general_frame} (or Section~\ref{sec:plm_example}) with $\widetilde{\eta}_{k}^{(j,m)}$ replaced by $\widehat{\eta}_{k}^{(j,m)}$.

The cross-validation procedure to select appropriate bandwidth $\hbar$ and tuning parameter $\widetilde{\lambda}$ can be implemented similarly. The $\hbar$ and $\widetilde{\lambda}$ can be selected to minimize the task-specific loss: \[\frac{1}{|M_m^j|}\sum_{i \in M_m^j}\left(Y_{k,i}-\widehat{\mu}_k^{(j,m)}(X_{k,i})\right)^2.\] Unlike the tuning procedure for $\lambda$, where a common $\lambda$ is chosen for all tasks, in tuning $\widetilde{\lambda}$ and $\hbar$, appropriate $\widetilde{\lambda}$ and $\hbar$ may be task-specific; this is a conclusion from Theorem~\ref{thm:plm}.

\begin{remark}[Higher-order kernel functions]
    Our estimation procedure in this section uses kernel functions of order $2$; higher-order kernel functions can also be adopted. Higher-order kernel functions can improve the bias-variance tradeoff in certain cases; however, they often have negative values in some regions, resulting in unsatisfied estimators, e.g., estimated propensity scores with negative values.
\end{remark}

\section{Theoretical results}
\label{sec:theory}

In this section, we present our main theoretical results along with their applications to specific examples.

\subsection{Main results}

To characterize the theoretical properties of the proposed methods, we define the relativeness among tasks based on parametric components and the nuisance relativeness based on nuisance parameters.
\begin{assn}(Task similarity)\label{assn:task_similarity}
    There exists a subset $S \subseteq [K]$ and $\theta_0$ such that $|S^c| \le \varepsilon K$ and $\max_{k \in S} \left\|\theta_k-\theta_0\right\|_2 \le \delta$.
\end{assn}
This assumption is denoted as $(\varepsilon,\delta)$-task relatedness throughout the paper. The smaller value of $\delta$ encourages similar tasks, and the parameter $\varepsilon$ bounds the number of outlier tasks. When $\varepsilon= \delta=0$, all the tasks are identical, and our method provides distributed learning for parameters of interest. In practice, $\varepsilon, \delta$ are unknown, and the proposed late fusion multi-task learning for the parametric components is expected to be robust against tasks with large deviations from the majority of tasks. This notion of task-relatedness was used by~\citet{duan2023adaptive,tian2023learning} for parametric models.

Now, we state the orthogonality condition of the estimating equation $m(\cdot)$ defined by \eqref{eq:define_score}. First, we assume true $\theta_k$'s satisfies the moment condition 
\begin{equation}\label{eq:moment_condition}
    \mathbb E_k \left[m(Z, \theta_k, \eta_k)\right]=0, \quad k \in [K],
\end{equation}
where $\eta_k \in \Lambda$ is the true nuisance parameter for $k$-th population, and $\Lambda$ is a convex subset of some normed vector space. Defining $\widetilde{\Lambda} = \{\eta -\eta_k, k \in [K], \eta \in \Lambda\}$, we denote the Gateaux derivative map by $D_{k,r}[\eta-\eta_k]= \partial_r E_k\left[m(Z, \theta_k, \eta_k + r(\eta-\eta_k))\right]$, $\eta \in \Lambda$. We require that the chosen estimating equation $m(Z, \theta_k, \eta_k)$ is Neyman near-orthogonal~\citet[Definition 2.2]{chernozhukov2018double}: There exists $\lambda_n= o(n^{-1/2})$ such that, $D_{k,r}[\eta-\eta_k]$ exists for all $r\in [0,1)$ and $\|D_{k,0}[\eta-\eta_k]\| \le \lambda_n$ for all $k \in [K]$.

Note that if $\lambda_n=0$, then $m(Z, \theta_k, \eta_k)$ is Neyman-orthogonal. Further, we require the following standard assumptions regarding Neyman near-orthogonality~\cite[Assumption 3.3]{chernozhukov2018double}:
\begin{assn}\label{assn:true_theta}
 The true parameters $\theta_k$ satisfies $\eqref{eq:define_score}$ and the parameter spaces contain a ball of radius $O(\log n /n)$ around $\theta_k$. The map $(\theta,\eta) \mapsto E_k\left[m(Z,\theta,\eta)\right]$ is twice Gateaux differentiable for all $(\theta,\eta)$. Further, there exists Jacobian matrices $J_k$ with eigenvalues between $c_0,c_1$ such that $2\|{E}_k\left[m(Z,\theta,\eta_k)\right]\| \ge \|J_k(\theta-\theta_k)\| \wedge c_0$. The score $m$ is $\lambda_n= \delta_n n^{-1/2}$ Neyman near-orthogonal.
 \end{assn}

This assumption is pretty mild and required to ensure that each $\theta_k$ is sufficiently separated from the boundary. The eigenvalue condition in Assumption~\ref{assn:true_theta} provides identifiability of the $\theta_k$'s. Next, we state our assumptions regarding regularity of score functions, along with estimation guarantees for the nuisance parameters.

\begin{assn}\label{assn:score}
   \begin{enumerate}
       \item[(i)] The nuisance parameter estimators belong to $\Lambda$ with probability $1-\Delta_n$. Moreover, the following rates hold: \begin{align*}
           &r_n= \max_{k \in [K]} \sup_{\eta,\theta} \|{E}\left[m(Z,\theta,\eta)\right]- {E}\left[ m(Z,\theta,\eta_k)\right]\| \le \delta_n \tau_n, \\
            &r'_n= \max_{k \in [K]} \sup_{\eta \in \Lambda, \|\theta-\theta_k\| < \tau_n} (E_k[\|m(Z,\theta,\eta)-m(Z,\theta_k,\eta_k)\|]^2)^{1/2},
       \end{align*} where we have $r'_n \log^{1/2}(1/r_n) \le \delta_n$. 
       \item[(ii)] All eigenvalues of the matrix $E_k\left[m(Z, \theta_k, \eta_k)m(Z, \theta_k, \eta_k)^\top\right]$ are bounded below by $c_0$ for all $k \in [K]$.
       \item[(iii)] The parameter spaces for $\theta_k$'s are uniformly bounded and $\mathcal{F}_\eta= \{m(.,\theta,\eta), \theta \in \Theta\}$ is measurable and uniform covering entropy satisfies $\max_{k \in [K]}\log N(\varepsilon \|\mathcal{F}_\eta\|_{Q,2}, F_{\eta}, \|.\|_{Q,2}) \le v \log(a/\varepsilon)$ for all $0< \varepsilon <1$, and some envelope function $F_{\eta}$.
       \end{enumerate}
\end{assn}
Here, the first assumption specifies mild growth conditions in terms of the quantities $\delta_n$ and $\tau_n$. The second condition posits that the variance of the scores is non-degenerate. Both Assumptions~\ref{assn:true_theta} and~\ref{assn:score} are common in semiparametric literature and we need them to have $\sqrt{n}$-consistency of our initial estimators $\tilde\theta_k$'s.


Equipped with these assumptions, we are now ready to state our first main theoretical result which characterizes the estimation rates for the optimization problem~\eqref{eq:multi-task_parametric}.
\begin{theorem}\label{thm:general}
    Suppose Assumptions~\ref{assn:task_similarity},~\ref{assn:true_theta} and~\ref{assn:score} hold and $\log K \ll n$. Further, there exists $\nu>0$ such that $\delta_n \le n^{-1/2+ 1/\nu} \log n$ and $n^{-1/2} \log n \le \tau_n \le \delta_n$. Suppose the initial estimators satisfy
    \begin{equation}\label{eq:theta_error}
        \max_{k \in [K]} \|\widetilde{\theta}_k-\theta_k\|^2_2 \le \delta^2_\theta =o_{\mathbb{P}}(1).
    \end{equation}
     Then $\widehat{\theta}_k$'s defined by \eqref{eq:multi-task_parametric} satisfies:
    \begin{equation}\label{eq:good_theta}
        \max_{k \in S} \|\widehat{\theta}_k-\theta_k\|_2= O_{\mathbb{P}}\Bigg(\frac{1}{\sqrt{nK}}+ \rho_n\Bigg)+\frac{6}{1-\varepsilon} \min\{3 \delta, \frac{2 \lambda}{5}\}+2 \varepsilon \lambda,
    \end{equation}
    where $\rho_n= r'_n\delta_{\theta}+\delta_{\theta}^2+ r_n$ and the tuning parameter $\lambda = C(\sqrt{\frac{\log K}{n}}+\rho_n)$. Moreover, for outlier tasks, we have that
    $
        \max_{k \in [K]}\|\widehat \theta_k -\theta_k\|_2 = O_{\P}\left(\sqrt{\frac{\log K}{n}}+\rho_n\right).
    $
\end{theorem}
Before proceeding further, we discuss the implications of Theorem~\ref{thm:plm}. Given the estimation error of the initial estimators $\widetilde{\theta}_k$ given in~\eqref{eq:theta_error}, the above result characterizes the error of the late-fusion estimator $\widehat{\theta}_k$. The leading term consists of both parametric and nonparametric error rates, denoted by $\frac{1}{\sqrt{nK}}$ and $\rho_n$, respectively. Further, the result illustrates the precise dependence of error bounds on the similarity of the tasks given by $(\varepsilon,\delta)$. Note that for $K=1$,  the use of a double-robust method yields a parametric rate since we obtain $n^{-1/2} > \rho_n$ for a broad class of examples. However, in the presence of many tasks, if the nonparametric components differ across tasks, each task incurs its own nonparametric error, making it impossible to achieve a parametric rate—even when all tasks share the same parametric component, i.e., $\theta_1= \ldots=\theta_k=\theta$. This result shows the necessity of multi-task learning for nuisance components to achieve a parametric rate, especially when $K$ is large. Theorem~\ref{thm:general} should be treated as a black-box estimation result since it does \textit{not} depend on how the initial estimators are obtained. The bound~\eqref{eq:good_theta} continues to hold for any off-the-shelf double machine learning estimator $\widetilde{\theta}_k$.

Note that the assumption $\delta^2_{\theta} = o_{\mathbb{P}}(1)$ in~\eqref{eq:theta_error} is fairly standard. If the initial estimators $\widetilde{\theta}_k$'s are sub-Gaussian with $\|\widetilde{\theta}_k - \theta_k\|_2 = O_{\mathbb{P}}(n^{-1/2})$, then using standard bounds for the maximum of sub-Gaussian variables, we obtain $\delta^2_{\theta} = (\log K)/n = o_{\mathbb{P}}(1)$, under the assumption that $\log K \ll n$. The complementary setting, where no consistent initial estimator of $\theta_k$ is available from individual tasks, is significantly more challenging and is left for future explorations.

\begin{remark}[Similarity between nuisance parameters]\label{rem:separate_nuisance}
    If the nuisance parameters $\eta_k$'s for the tasks $k \in S$ are related, one can hope to achieve better estimation guarantees for those $\eta_k$'s. This will result in the following improvement of Theorem \ref{thm:general}: Suppose the definition of $r_n$ in Assumption \ref{assn:score} is replaced by $$r_n= \max_{k \in S} \sup_{\eta,\theta} \|{E}\left[m(Z,\theta,\eta)\right]- {E}\left[ m(Z,\theta,\eta_k)\right]\|.$$
    The conclusion of Theorem \ref{thm:general} still holds with $\lambda= \sqrt{\frac{\log K}{n}}+ \rho_n$, where $\rho_n= r'_n\delta_{\theta}+\delta_{\theta}^2+ r_n$ with $r_n$ defined here. Since $r_n$ improves estimation rates only for the related tasks $k \in S$, similarity in the nuisance parameters can lead to substantial gains in estimation accuracy, as demonstrated in the subsequent theoretical results and numerical experiments. 
\end{remark}

Theorem~\ref{thm:general} provides the estimation error bounds for the parametric components. Next, we focus on the theoretical properties of multi-task learning for nuisance parameters as described in Section~\ref{sec:mtl_nonpar}. To this end, we introduce the concept of relatedness of nuisance parameters similar to Assumption \ref{assn:task_similarity}.  
\begin{assn}(Nuisance similarity)\label{assn:nuisance_similarity}
    There exists a subset $S_{\eta} \subseteq [K]$ and $\eta_0$ such that $$\max_{k \in S_{\eta}} \left\|\eta_k-\eta_0\right\| \le \delta_{\eta}$$ almost surely, and $|S_{\eta}^c| \le \varepsilon_{\eta} K$, where $\eta_k$'s is an infinite-dimensional nuisance parameter defined as a conditional mean, e.g., $E_k[Y_{k,i}\mid X_{k,i}]$. We assume that $\eta_k$'s are at least second-order differentiable with uniformly bounded gradients; the density functions of $X_{k,i}$'s are uniformly bounded away from $0$ and $+\infty$.
\end{assn}
This assumption is denoted as $(\varepsilon_{\eta},\delta_{\eta})$-nuisance relatedness. Similar to task relatedness, this assumption encourages similar nuisance parameters across tasks. The smoothness conditions imposed on $\eta_k$'s ar standard in Kernel regression literature. To the best of our knowledge, our work is the first to consider a semiparametric framework based on Assumptions~\ref{assn:task_similarity} and~\ref{assn:nuisance_similarity}. Under Assumption~\ref{assn:nuisance_similarity}, we have the following theorem.
\begin{theorem}\label{thm:mtl_nuisance}
    Fix and $t \ge 2$, set $\widetilde{\lambda}= \left(\sqrt{\frac{\log K\hbar^{-2p}}{n\hbar^{p}}}+\hbar^2\right)$. Suppose the estimators of $\eta_k$'s are obtained following the procedure described in Section~\ref{sec:mtl_nonpar}, then $\widehat{\eta}_k$'s satisfy
    \begin{equation*}
        \max_{k \in S_{\eta}} \|\widehat{\eta}_k-\eta_k\| \le O_{\P} \left( \sqrt{\frac{\log (K\hbar^{-2p})}{nK\hbar^{p}}}+\hbar^2 +\min \Big\{\delta_{\eta}, \sqrt{\frac{\log (K\hbar^{-2p})}{n\hbar^{p}}}\Big\} +\varepsilon_{\eta}\sqrt{\frac{\log K\hbar^{-2p}}{n\hbar^{p}}}\right).
    \end{equation*}
    Moreover, for outlier tasks, we have that
    $
        \max_{k \in [K]}\|\widehat{\eta}_k-\eta_k\| = O_{\P}\left(\sqrt{\frac{\log 
        K\hbar^{-2p}}{n\hbar^{p}}}+\hbar^2\right).
    $
\end{theorem}

Theorem~\ref{thm:mtl_nuisance} provides estimation guarantees for the nuisance parameters from the proposed method. Similar to Theorem~\ref{thm:general}, the Theorem shows an improvement of $K^{-1/2}$ if $k \in S_\eta$. Further, the above result implies that in the natural case where $S= S_\eta$ where the there is a good set of tasks which share both the parametric and nonparametric components, plugging in the estimator $\hat\eta_k$ in Algorithm~\ref{algo:general} leads to faster convergence rate of learning. 

\subsection{Example: partial linear model} 

Theorems~\ref{thm:general} and~\ref{thm:mtl_nuisance} provides estimation rates for multi-task learning for parametric and nonparametric components, respectively. In this section, we focus on the specific example of partial linear models given by Example~\ref{example:1} and derive the error bounds. Theoretical results for Example~\ref{example:2} and~\ref{example:3} can be found in the online Supplementary Material. For the partial linear model (see Example~\ref{example:1}), its implementation follows Sections~\ref{sec:plm_example} and~\ref{sec:mtl_nonpar}. Notice that the initial estimator is constructed using cross-fitting based on a doubly robust estimating equation. Based on the specific construction of the initial estimator, by Theorems~\ref{thm:general} and~\ref{thm:mtl_nuisance}, we have the following theorem.

\begin{theorem}[Partial Linear Model]\label{thm:plm}
    Set $\lambda=(nK)^{-1/2}+\widetilde{a}_n^2+ \widetilde{a}_n\delta_{\theta}+\delta_{\theta}^2$ and $\widetilde{\lambda}= \sqrt{\frac{\log K\hbar^{-2p}}{n\hbar^{p}}}+\hbar^2$, where $\widehat{a}_n=\sqrt{\frac{\log K\hbar^{-2p}}{nK\hbar^{p}}}+\hbar^2 +\min \Big\{\delta_{\eta}, \sqrt{\frac{\log K\hbar^{-2p}}{n\hbar^{p}}}\Big\} +\varepsilon_{\eta}\sqrt{\frac{\log K\hbar^{-2p}}{n\hbar^{p}}}$ and $\widetilde{a}_n=\sqrt{\frac{\log 
        K\hbar^{-2p}}{n\hbar^{p}}}+\hbar^2$.
    \begin{enumerate}
        \item For $k\in S\cap S_{\eta}$, we have
        \begin{equation*}
        \|\widehat{\theta}_k-\theta_k\| \le O_{\P} \left( \sqrt{\frac{\log K}{nK}}+\widehat{a}_n^2 +\min \Big\{\delta, \lambda\Big\} +(\varepsilon+\varepsilon_{\eta})\lambda\right);
    \end{equation*}
     \item For $k\in S/ S_{\eta}$, we have
        \begin{equation*}
        \|\widehat{\theta}_k-\theta_k\| \le O_{\P} \left( \sqrt{\frac{\log K}{nK}}+\widetilde{a}_n^2 +\min \Big\{\delta, \lambda\Big\} +\varepsilon\lambda\right);
    \end{equation*}
    \item For $k\not\in S$, we have that
        $
        \|\widehat{\theta}_k-\theta_k\| \le O_{\P}\left(\sqrt{\frac{\log K}{n}}+\widetilde{a}_n^2\right),
    $
    \end{enumerate}
    where $\widehat{\theta}_k$'s are defined by procedures in Section~\ref{sec:plm_example} with $\eta_k$'s estimated using procedures in Section~\ref{sec:mtl_nonpar}.
\end{theorem}
Theorem~\ref{thm:plm} implies that tasks can be separated into three types with different error rates based on task similarity. For the tasks with both similar $\theta_k$ and $\eta_k$ (i.e., $k\in S\cap S_{\eta}$), naturally the resulting estimator can benefit from multi-task learning for both parametric and nuisance components, as reflected by the quantities $\sqrt{\frac{\log K}{nK}}$ and $\widehat{a}_n^2$ respectively. In contrast, if the nonparametric components are not shared, i.e., for $k\in S/ S_{\eta}$, only the benefit from multi-task learning for parametric components shows up. In this case, the error bound includes $\widetilde{a}_n^2$, task-specific error bound for $\eta_k$'s. Finally, for the \textit{bad} tasks $k\not\in S$, Theorem~\ref{thm:plm} shows robustness of our method.

\begin{remark}[Tuning parameters]\label{remark:4}
    Theorem~\ref{thm:plm} implies that the tuning of $(\widetilde{\lambda},\hbar)$ may need a different strategy compared with the tuning procedure described in Remark~\ref{remark:tuning} when nuisance similarity is considered to improve estimation. When nuisance similarity is considered, Theorem~\ref{thm:plm} shows that the optimal choice of $\hbar$ and thus $\widetilde{\lambda}$ (that minimizes the convergence rate) depends on the specific type of the tasks, which is unknown. Thus, in practice, we can select task-specific $(\widetilde{\lambda},\hbar)$ that minimizes the task-specific loss for tuning $\widetilde{\lambda}$ and $\hbar$.
\end{remark}

The main takeaway from Theorem \ref{thm:plm} is to show how kernel regression can be useful to exploit similarity among nuisances and obtain better estimation. One might wonder if the error rates provided by Theorem \ref{thm:plm} can be achieved by off-the-shelf ML algorithms. In the next result, we characterize the rate if one chooses the initial estimators $\tilde \theta_k$'s using double machine learning estimators and the nuisance parameters $f_k$ and $g_k$ be estimated independently using any machine learning method. Note that, owing to the efficiency of the double machine learning estimators, one can use the simplified optimization instead of \eqref{eq:multi-task_parametric}: 
\begin{equation}\label{eq:quad_opt}
    \min_{u_0, u_1, \ldots u_K} \sum_{k=1}^{K} \left\{V_k\|\tilde \theta_k-u_k\|^2_2+ \lambda \|u_k -u_0\|_2\right\},
\end{equation}
where $V_k$ is the inverse of estimated variance of $\tilde \theta_k$'s (Details provided in online Supplementary Material) and the minimizer $u_k$ is denoted by $\hat \theta_k$, $k=1,\ldots, K$. We establish the following estimation result:

\begin{prop}\label{prop:plm_dml}
    Suppose the estimators $\hat f_k$ and $\hat g_k$ of $f_k$ and $g_k$ satisfies $\sup_{k \in K}\|\hat f_k -f_k\|_2= O(n^{-\alpha})$, $\sup_{k \in K}\|\hat g_k -g_k\|_2= O(n^{-\beta})$. Set $\lambda= C\sqrt{\frac{\log K+t}{n}}$ for $t \ge 2$. Then we have 
    \begin{equation*}
        \max_{k\in S} \|\hat \theta_k - \theta_k\|_2 =O_{\P} \left( \frac{1}{\sqrt{nK}}+n^{-\alpha-\beta}+\frac{6}{1-\varepsilon} \min\{3 \delta, \frac{2 \lambda}{5}\}+2 \varepsilon \lambda \right).
    \end{equation*}
Moreover, we have $\max_{k\notin S} \|\hat \theta_k - \theta_k\|_2 =O_{\P} \left( \sqrt{\frac{\log K}{n}}\right)$.
\end{prop}

Proposition \ref{prop:plm_dml} precisely characterizes the bias-variance trade-off for multi-task learning in partial linear models using any black-box machine learning method. The error bounds, $\alpha$ and $\beta$ are problem-specific. For general $\gamma$-smooth functions of nuisance parameters, these bounds are given by $\alpha=\beta= \frac{\gamma}{2 \gamma+p}$~\cite{kohler2005adaptive} when deep neural networks are used for estimation. Subsequent works have extended established estimation guarantees using deep neural networks for broader function classes, including the generalized hierarchical interaction model~\cite{bauer2019deep,kohler2016nonparametric,kohler2021rate,schmidt2020nonparametric}. The proof of Proposition \ref{prop:plm_dml} is provided in the online Supplementary Material. Finally, it is an practical concern how to obtain improve estimation of nuisance parameters when they are similar using ML methods --- we believe this direction requires significantly new ideas and we leave it for future explorations.

\section{Simulations}
\label{sec:simulation}

In this section, we conduct simulations to demonstrate the efficacy of our estimator. We compare the errors of our proposed late fusion multi-task learning estimator with 
individual task learning. Moreover, we illustrate how the estimations of nuisance parameters affect the performance of the resulting estimator. To this end, we consider three scenarios corresponding to a partial linear model, a single-index model, and a single-index model for the CATE. For each setup, we generate five tasks with four similar tasks and one outlier. For each task, we independently generate $8$ covariates following a uniform distribution on $[-1,1]$. Here are the details of each scenario:
\begin{enumerate}
\item\label{sim:CATE} We generate $Y_{k,i}$ following a Bernoulli distribution with a success probability of $$\textrm{logit}\left\{T_{k,i}f_k(X_{k,i}^\top \theta_k)+\mu_k(X_{k,i})\right\}.$$ To generate similar tasks (i.e., for $k\leq 4$), we follow $f_k(X_{k,i}^\top \theta_k)=X_{k,i}^\top\theta_{k}$, where $\theta_k=\theta_0+0.01b_k$, $b_k$'s follow standard normal distributions, and $\theta_0=(-1,1,0.5,0.5,-0.5,-0.5,0,0)^\top$; to generate the outlier task, we follow $\theta_{5}=\theta_0+0.01b_k+0.5(1-\eta)$. We set $\mu_k(X_{k,i}^\top\theta_{k})=0.2\times \left(X_{k,i}^\top\theta_{k}+1\right)^2$. We generate $T_{k,i}$ from a Bernoulli distribution with a success probability of $\textrm{logit}\left\{4\pi_k(X_{k,i})-1\right\}$. We set $\pi_k(X_{k,i})=X_{k,i}^{(1)}$ for $k\leq 3$ and $\pi_k(X_{k,i})=X_{k,i}^{(1)}+0.5\left\{X_{k,i}^{(2)}\right\}^2$ for others.

    \item\label{sim:sim} We generate $Y_{k,i}$ following a single-index model: $Y_{k,i}=f_k(X_{k,i}^\top\theta_k)+\epsilon_{k,i},$ where $\epsilon_{k,i}$ follows a normal distribution with mean $0$ and a standard deviation of $0.2$. To generate similar tasks (i.e., for $k\leq 4$), we follow $f_k(X_{k,i}^\top \theta_k)=X_{k,i}^\top\theta_{k}$, where $\theta_k=\theta_0+0.05b_k$, $b_k$'s follow standard normal distributions, and $\theta_0=(2,-2,1,-1,0,0,0,0)^\top$; to generate the outlier task, we follow $\theta_{5}=\theta_0+0.05b_k+(1-\eta)$. We set $f_k(X_{k,i}^\top\theta_k)=0.2\times \left(X_{k,i}^\top\theta_{k}\right)^2$.

    \item\label{sim:plm} We generate $Y_{k,i}$ following a partial linear model: $Y_{k,i}=T_{k,i}\theta_k+f_k(X_{k,i})+\epsilon_{k,i},$ where $\epsilon_{k,i}$ follows a standard normal distribution. To generate similar tasks (i.e., for $k\leq 4$), we follow $\theta_k=2+0.1(G_k+1)(1-\eta)$, where $G_k$ follow a standard normal distribution; to generate the outlier task, we follow $\theta_5=2+0.1(G_5+1)(1-\eta)+0.1(1-\eta)$, where $\eta$ is a parameter controlling the similarity between the fifth task and others. We set $f_k(X_{k,i})=0.5\left\{\sum_{l=1}^8{X}_{k,i}^{(l)}\right\}^2+\left\{\sum_{l=1}^8{X}_{k,i}^{(l)}\right\}^3$, where $X_{k,i}^{(l)}$ $l$-th covariate of $X_{k,i}$. To generate $T_{k,i}$, we follow $T_{k,i}=g_k(X_{k,i})+\widetilde{\epsilon}_{k,i}$, where $\widetilde{\epsilon}_{k,i}$ follows a normal distribution with mean $0$ and a standard deviation of $0.5$. We set $g_k(X_{k,i})=0.25\times \left(\left\{X_{k,i}^{(1)}\right\}^2+\left\{X_{k,i}^{(2)}\right\}^2\right)$ for $k \leq 3$ and $g_k(X_{k,i})=0.25\times \left(\left\{X_{k,i}^{(1)}\right\}^2+\left\{X_{k,i}^{(2)}\right\}^2\right)+G_k(1-\eta)\left(\left\{X_{k,i}^{(3)}\right\}^2+\left\{X_{k,i}^{(4)}\right\}^2-2\right)$ for others.

\end{enumerate}
In Scenarios~\ref{sim:CATE} -~\ref{sim:plm}, we incorporate a simulation parameter $\eta$, which controls the similarity between the outlier task and other tasks. When $\eta=1$, the parameters of interest are similar (up to a small perturbation) across five tasks; when $\eta=0$, the parameters of interest are similar among Tasks~1 -~4 but different in Task~5. Overall, among these five tasks, Tasks~1 -~4 always have similar parametric components, i.e., $\theta_k$'s. Although Task 4 has similar parametric components, the specification of nuisance parameters may differ from others, e.g., the propensity score $\pi_k(X_{k,i})$ in Scenario~\ref{sim:CATE}. Task 5 is the outlier task, which has different specifications in both parametric components and nuisance parameters. In our simulation settings, we vary $\eta$ from $0$ to $1$ by $0.1$, and sample sizes of each task, i.e., $100$ and $200$, leading to a total $22$ simulation settings for each scenario. Under these different simulation settings, we can thoroughly examine the performance of the proposed multi-task learning with and without late fusion for nuisance parameters. To further compare with other methods, we implement the Oracle Trans-Lasso proposed in \citet{li2022transfer} with a linear model including both covariate and treatment interactions to estimate CATE. In this implementation, we treat each task as a target task and others as source tasks.

For each simulation setting, we generate the training dataset for each task and implement each method on the generated training datasets. For our proposed method, when implementing, we simulate five local servers (i.e., one server is used for one task) and one central server and ensure no direct raw data sharing between local servers and the central server. From the central server, we obtain the parameter estimates for each task. When implementing individual task learning, we directly use the training data in each local server to estimate the parameters. Then, we calculate the mean squared error (MSE) for the parameters in each task. We report the averaged MSE, individual and the maximum MSEs across five different tasks. This procedure, including training data generation, method implementation, and MSE calculations, is repeated  $100$ times.

\begin{figure}
\begin{center}
	\includegraphics[width=\linewidth]{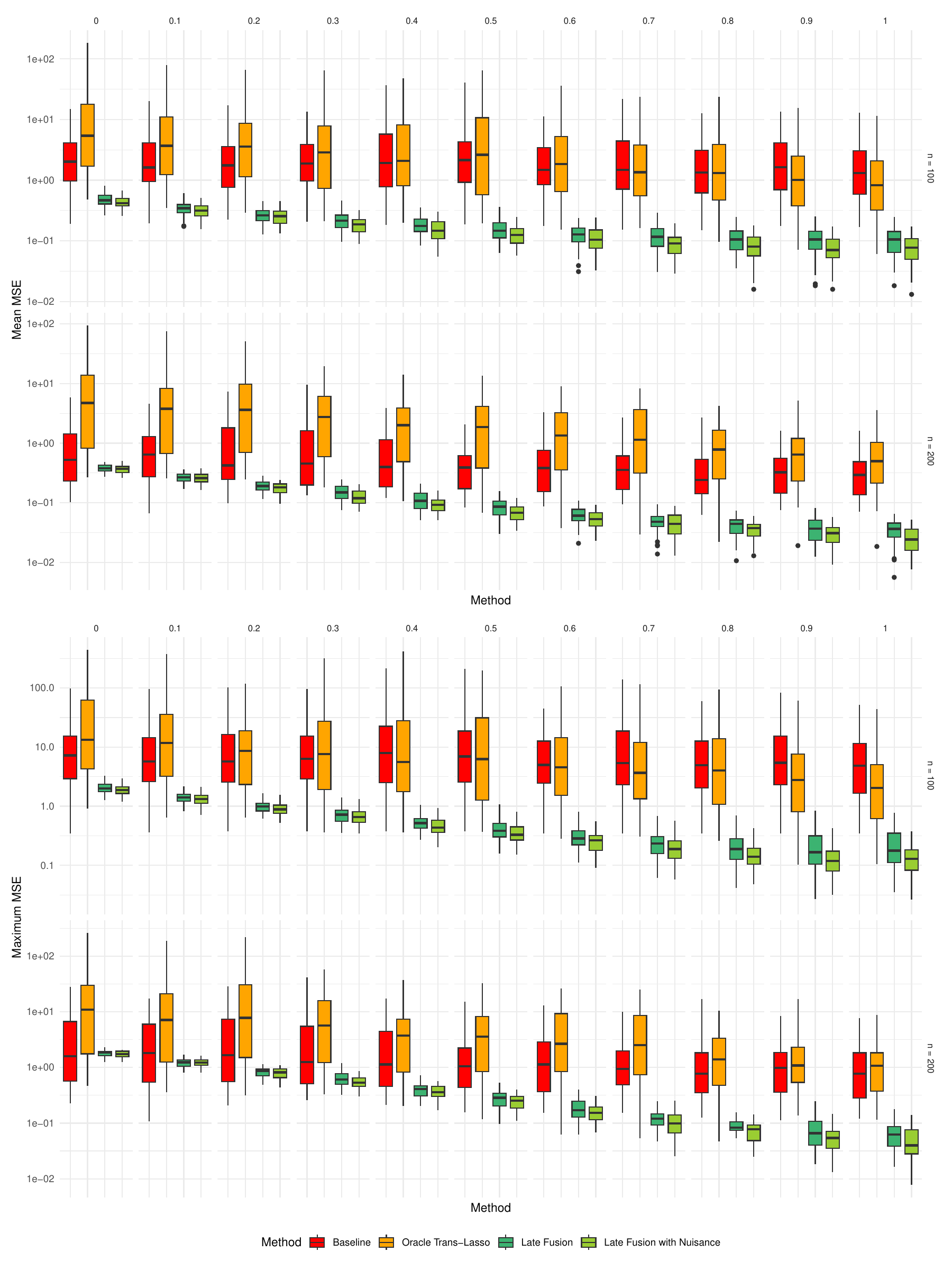}
\end{center}
	\caption{Averaged and maximum MSE for Scenario~\ref{sim:CATE} over $100$ repeats. The x-axis represents different values of $\eta$. \label{fig:cate1}}
\end{figure}

\begin{figure}
    \centering
    \includegraphics[width=\linewidth]{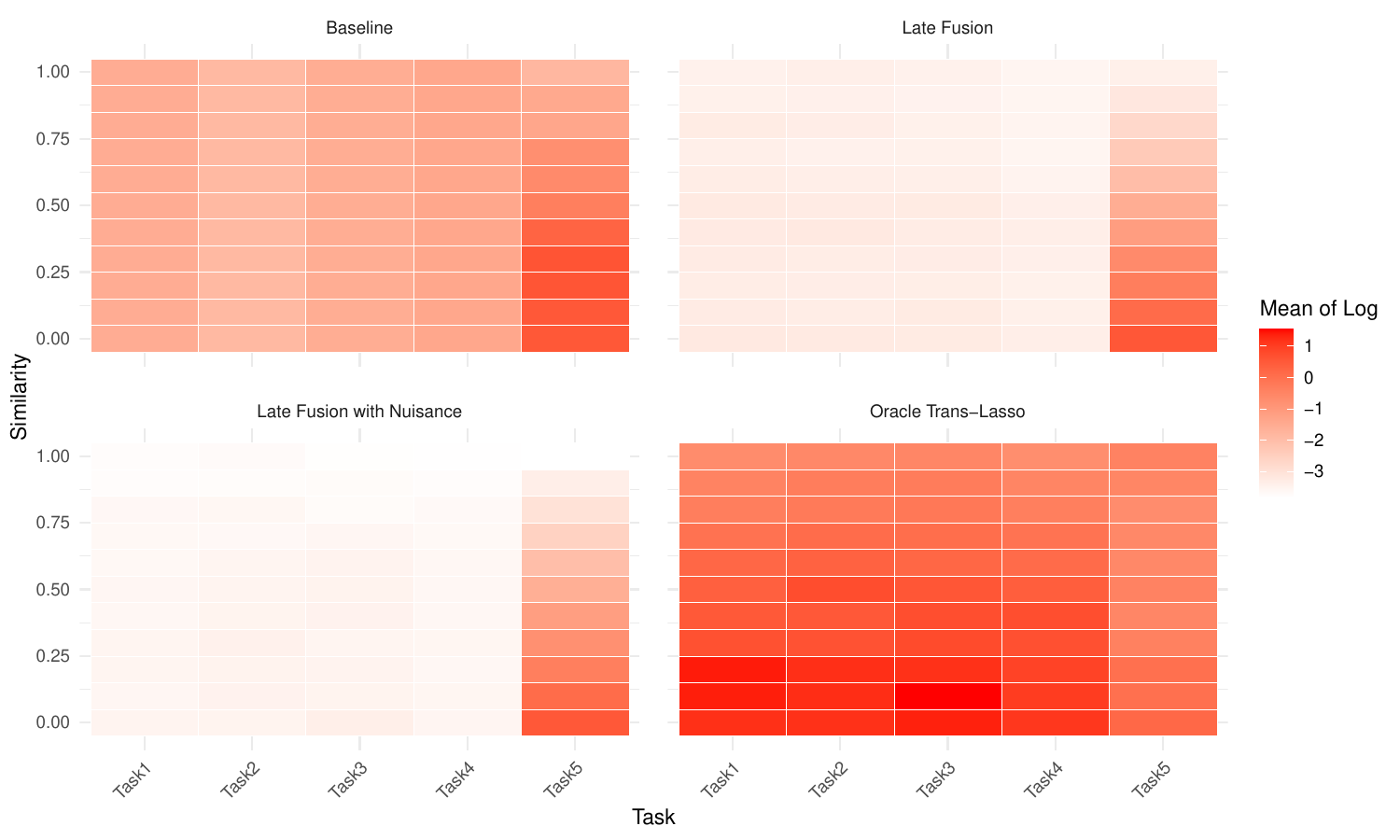}
    \caption{Log-transformed averaged MSE for each task in Scenario~1 over 100 repeats. In each subplot, the y-axis represents different values of $\eta$.}
    \label{fig:heat_CATE}
\end{figure}

Figures~\ref{fig:cate1} and~\ref{fig:heat_CATE} summarize the results in Scenario~\ref{sim:CATE}. The imulation results for the scenarios~\ref{sim:sim} and~\ref{sim:plm} can be found in the online Supplementary Material. From Figure~\ref{fig:cate1}, our proposed multi-task learning methods achieve lower averaged MSE and maximum MSE than the individual-task learning and Oracle Trans-Lasso across all different similarities among five tasks. When $\eta$ is close to $1$, i.e., all five tasks are similar, the advantages of our proposed methods compared to the individual-task learning are larger. Although Oracle Trans-Lasso has a misspecified model, it outperforms individual-task learning when the sample size is very limited and $\eta$ is close to $1$ due to a better bias and variance trade-off. Further, our multi-task learning with late fusion for nuisance parameters achieves the lowest averaged MSE and maximum MSE. This shows the advantages of adopting late fusion multi-task learning for nuisance parameters, as suggested in Theorem~\ref{thm:mtl_nuisance}. Figure~\ref{fig:heat_CATE} summarizes the performance for each task under varying values of $\eta$ when sample size is fixed at 200. For Task 1 to Task 4, our methods consistently achieve the lowest average MSE and remain stable across different similarity levels. Task 5 exhibits comparable MSE across all methods, which is consistent with our theoretical findings, especially when $\eta$ (i.e., similarity) is small. Overall, our proposed methods can achieve robust performance gains compared with individual task learning and provide a more flexible alternative to Oracle Trans-Lasso. The late fusion for nuisance parameters can lead to better performance. 

\section{Application to Mammography Screening Study}
\label{sec:real_data}

In this section, we apply our proposed method to estimate the CATE of phone consulting in mammography screening adherence. Our datasets include two randomized controlled trials: Computer And Phone (CAPE) study and the National Institute of Nursing Research (NINR) study. Both studies focus on the efficacy of interventions to promote mammography screening. One common intervention in the two trials is phone consultation. The published findings of the NINR and CAPE studies identified possible treatment effect modifiers, including education, marital status, and beliefs. In our study, we apply our proposed method to estimate CATEs using two studies without the need for direct data sharing. The outcome of interest is whether the patients underwent mammography screening $21$ months after the baseline. The covariates include socioeconomic status (working for pay or not), race (Caucasian or not), marital status, baseline stage of mammography screening behavior, number of years had a mammogram in the past 2 to 5 years, family history of cancer, and several belief and knowledge scale scores. A descriptive table (e.g., sample sizes and variable distribution comparisons) on the population of two studies can be found in the online Supplementary Material. It shows that the NINR had older women with lower barrier scale scores compared with the CAPE data set. Data from different trials are considered different tasks; we also split each trial into two datasets (treated as two tasks per trial) to reduce the computational complexity. In total, we have four tasks that are expected to be similar. We assume a single-index model for CATE, which shares the same setups as in Example~\ref{example:3}.

\begin{figure}[ht]
\begin{center}
	\includegraphics[width=\linewidth]{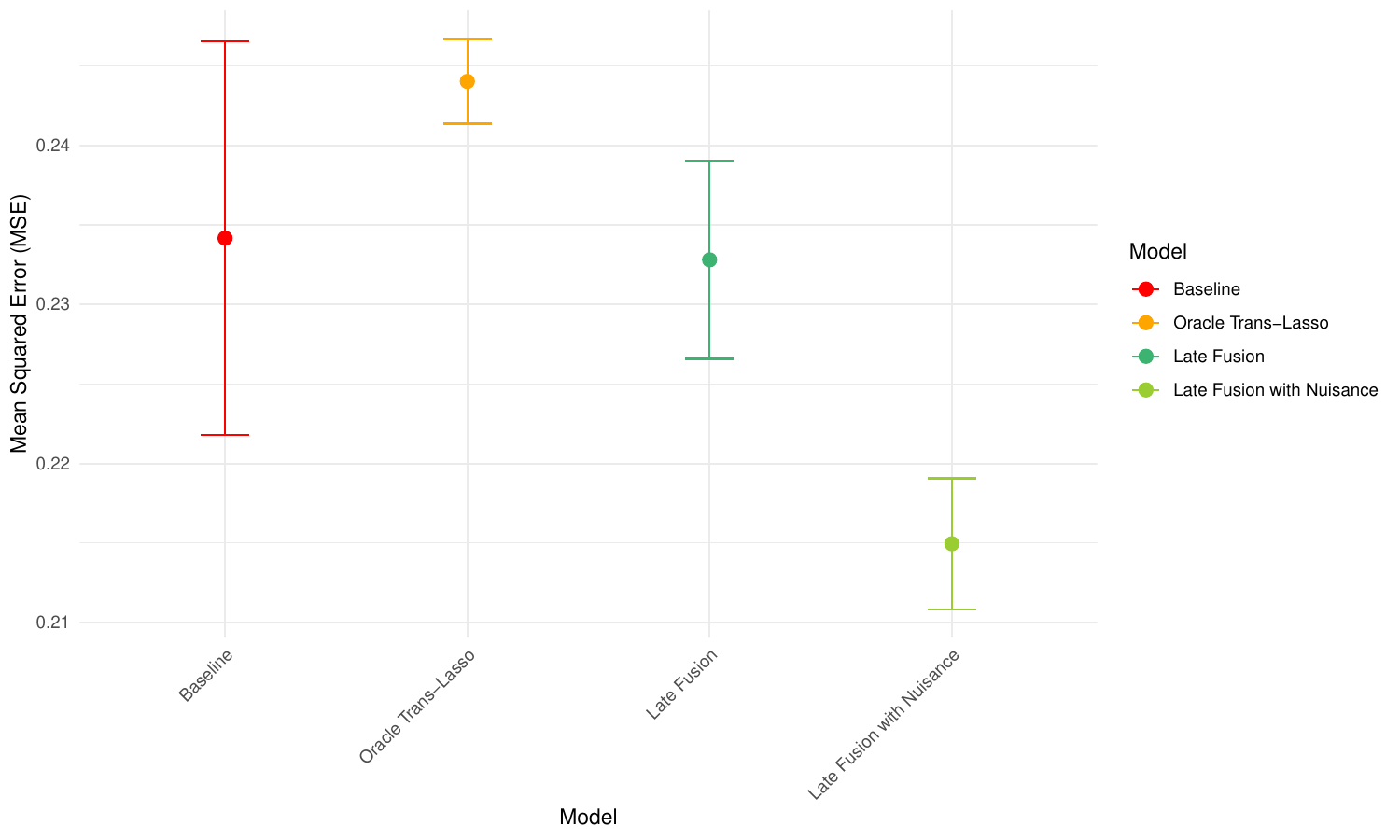}
\end{center}
	\caption{Averaged and Maximum MSE for the application to mammography screening study over $100$ repeats. The dot is the mean of the averaged MSE over 100 repeats; the upper and lower bar is the 95\%-CI of the mean. \label{fig:real_data1}}
\end{figure}

We conduct two analyses using the proposed methods. In the first analysis, we split each task into training and testing datasets with a sample size ratio of 3:2. In the training datasets, we implement the proposed methods, including the proposed methods with and without late fusion for nuisance parameters, the Oracle Trans-Lasso, and individual task learning. On the testing datasets, we calculate the prediction error following
           \begin{eqnarray*}
    \widehat{E}_{\text{test}}\left[\frac{1}{\widehat{\pi}_k(T, X)}\left(Y-\frac{1}{2}T\widehat{f}_k (X^\top\widehat{\theta}_k)-\widehat{\mu}_k(X)\right)^2\right],
\end{eqnarray*}
where $\widehat{\theta}_k$ is the estimates using different methods, $\widehat{f}_k(\cdot)$ is estimated using data in each task, and $\widehat{E}_{\text{test}}[\cdot]$ represents the empirical average over the testing dataset. The $\widehat{\pi}_k(\cdot, \cdot)$ and $\widehat{\mu}_k(\cdot)$ as nuisance parameters are estimated via kernel regressions. In addition, the nuisance parameters and the estimation of prediction error are implemented following the cross-fitting algorithm. The entire procedure, including splitting the data sets, implementing methods on training data sets, and calculating the prediction errors on the testing data sets, is repeated $100$ times. We report the averaged and maximum prediction errors over four tasks. Figure~\ref{fig:real_data1} shows that the averaged prediction errors over $100$ repeat. The proposed method with late fusion for nuisance parameters achieves the lowest prediction errors; the proposed late fusion method for only parametric components performs comparable to the baseline, i.e. individual task learning. This indicates that the estimation of nuisance parameters contributes significantly to the prediction errors; our proposed method can leverage the nuisance similarity to improve the estimation of nuisance parameters and thus, each task. In the second analysis, we apply the proposed multi-task learning with late fusion for nuisance parameters on the entire dataset. Then we calculate the correlation of each covariate with the predicted CATE to interpret the relationship between the predicted direction that dominates the CATEs. Details can be found in the online Supplementary Material.

\section{Discussion}
\label{sec:diss}

Our work addresses the critical challenge of developing rigorous and computationally efficient algorithms for multi-task learning in semiparametric problems with infinite-dimensional nuisance parameters. The proposed late fusion framework is privacy-preserving and robust against outlier tasks. It follows a two-step approach: first, initial estimators are derived from individual task learning by solving estimating equations; second, these estimators are adaptively aggregated to leverage similarities across tasks while maintaining robustness. Importantly, our methodology does not rely on assumptions about the similarity of nuisance parameters between tasks. Nevertheless, for scenarios where such similarity exists, we introduce an innovative late fusion multi-task nonparametric learning approach for nuisance parameters. This approach provides improved estimation guarantees for nuisance parameters, which ultimately improve the estimation of parameters of interest.

This work lays the foundation for several promising avenues of future research. One direction is to extend the proposed framework to settings where the i.i.d. assumption does not hold, such as multi-task learning for longitudinal data analysis. Addressing dependence structures in the data would significantly broaden the applicability of the method. Another compelling direction is to incorporate alternative forms of task similarity, such as clustering or low-rank structures. For instance, if an underlying clustering structure exists among multiple treatments, data-driven multi-task clustering approaches could be developed to estimate CATEs for each treatment more efficiently.

\bibliography{ref}

\begin{thebibliography}{}

\bibitem[Baltru{\v{s}}aitis et~al., 2018]{baltruvsaitis2018multimodal}
Baltru{\v{s}}aitis, T., Ahuja, C., and Morency, L.-P. (2018).
\newblock Multimodal machine learning: A survey and taxonomy.
\newblock {\em IEEE transactions on pattern analysis and machine intelligence},
  41(2):423--443.

\bibitem[Bastani, 2021]{bastani2021predicting}
Bastani, H. (2021).
\newblock Predicting with proxies: Transfer learning in high dimension.
\newblock {\em Management Science}, 67(5):2964--2984.

\bibitem[Bauer and Kohler, 2019]{bauer2019deep}
Bauer, B. and Kohler, M. (2019).
\newblock On deep learning as a remedy for the curse of dimensionality in
  nonparametric regression.
\newblock {\em The Annals of Statistics}, 47(4):2261.

\bibitem[Bodory et~al., 2022]{bodory2022evaluating}
Bodory, H., Huber, M., and Laff{\'e}rs, L. (2022).
\newblock Evaluating (weighted) dynamic treatment effects by double machine
  learning.
\newblock {\em The Econometrics Journal}, 25(3):628--648.

\bibitem[Bonvini et~al., 2024]{bonvini2024doubly}
Bonvini, M., Kennedy, E.~H., Dukes, O., and Balakrishnan, S. (2024).
\newblock Doubly-robust inference and optimality in structure-agnostic models
  with smoothness.
\newblock {\em arXiv preprint arXiv:2405.08525}.

\bibitem[Cai and Pu, 2024]{cai2024transfer}
Cai, T.~T. and Pu, H. (2024).
\newblock Transfer learning for nonparametric regression: Non-asymptotic
  minimax analysis and adaptive procedure.
\newblock {\em arXiv preprint arXiv:2401.12272}.

\bibitem[Chen et~al., 2011]{chen2011integrating}
Chen, J., Zhou, J., and Ye, J. (2011).
\newblock Integrating low-rank and group-sparse structures for robust
  multi-task learning.
\newblock In {\em Proceedings of the 17th ACM SIGKDD international conference
  on Knowledge discovery and data mining}, pages 42--50.

\bibitem[Chernozhukov et~al., 2018]{chernozhukov2018double}
Chernozhukov, V., Chetverikov, D., Demirer, M., Duflo, E., Hansen, C., Newey,
  W., and Robins, J. (2018).
\newblock Double/debiased machine learning for treatment and structural
  parameters: Double/debiased machine learning.
\newblock {\em The Econometrics Journal}, 21(1).

\bibitem[Colangelo and Lee, 2020]{colangelo2020double}
Colangelo, K. and Lee, Y.-Y. (2020).
\newblock Double debiased machine learning nonparametric inference with
  continuous treatments.
\newblock {\em arXiv preprint arXiv:2004.03036}.

\bibitem[D{\'\i}az, 2020]{diaz2020machine}
D{\'\i}az, I. (2020).
\newblock Machine learning in the estimation of causal effects: targeted
  minimum loss-based estimation and double/debiased machine learning.
\newblock {\em Biostatistics}, 21(2):353--358.

\bibitem[Duan et~al., 2022]{duan2022heterogeneity}
Duan, R., Ning, Y., and Chen, Y. (2022).
\newblock Heterogeneity-aware and communication-efficient distributed
  statistical inference.
\newblock {\em Biometrika}, 109(1):67--83.

\bibitem[Duan and Wang, 2023]{duan2023adaptive}
Duan, Y. and Wang, K. (2023).
\newblock Adaptive and robust multi-task learning.
\newblock {\em The Annals of Statistics}, 51(5):2015--2039.

\bibitem[Dukes et~al., 2020]{dukes2020doubly}
Dukes, O., Avagyan, V., and Vansteelandt, S. (2020).
\newblock Doubly robust tests of exposure effects under high-dimensional
  confounding.
\newblock {\em Biometrics}, 76(4):1190--1200.

\bibitem[Dukes and Vansteelandt, 2021]{dukes2021inference}
Dukes, O. and Vansteelandt, S. (2021).
\newblock Inference for treatment effect parameters in potentially misspecified
  high-dimensional models.
\newblock {\em Biometrika}, 108(2):321--334.

\bibitem[Feuerriegel et~al., 2024]{feuerriegel2024causal}
Feuerriegel, S., Frauen, D., Melnychuk, V., Schweisthal, J., Hess, K., Curth,
  A., Bauer, S., Kilbertus, N., Kohane, I.~S., and van~der Schaar, M. (2024).
\newblock Causal machine learning for predicting treatment outcomes.
\newblock {\em Nature Medicine}, 30(4):958--968.

\bibitem[Gu et~al., 2022]{gu2022robust}
Gu, T., Han, Y., and Duan, R. (2022).
\newblock Robust angle-based transfer learning in high dimensions.
\newblock {\em arXiv preprint arXiv:2210.12759}.

\bibitem[Heiler and Knaus, 2021]{heiler2021effect}
Heiler, P. and Knaus, M.~C. (2021).
\newblock Effect or treatment heterogeneity? policy evaluation with aggregated
  and disaggregated treatments.
\newblock {\em arXiv preprint arXiv:2110.01427}.

\bibitem[Hitsch et~al., 2024]{hitsch2024heterogeneous}
Hitsch, G.~J., Misra, S., and Zhang, W.~W. (2024).
\newblock Heterogeneous treatment effects and optimal targeting policy
  evaluation.
\newblock {\em Quantitative Marketing and Economics}, 22(2):115--168.

\bibitem[Hospedales et~al., 2021]{hospedales2021meta}
Hospedales, T., Antoniou, A., Micaelli, P., and Storkey, A. (2021).
\newblock Meta-learning in neural networks: A survey.
\newblock {\em IEEE transactions on pattern analysis and machine intelligence},
  44(9):5149--5169.

\bibitem[Hunter and Holmes, 2023]{hunter2023medical}
Hunter, D.~J. and Holmes, C. (2023).
\newblock Where medical statistics meets artificial intelligence.
\newblock {\em New England Journal of Medicine}, 389(13):1211--1219.

\bibitem[Jalali et~al., 2013]{jalali2013dirty}
Jalali, A., Ravikumar, P., and Sanghavi, S. (2013).
\newblock A dirty model for multiple sparse regression.
\newblock {\em IEEE Transactions on Information Theory}, 59(12):7947--7968.

\bibitem[Jordan et~al., 2019]{jordan2019communication}
Jordan, M.~I., Lee, J.~D., and Yang, Y. (2019).
\newblock Communication-efficient distributed statistical inference.
\newblock {\em Journal of the American Statistical Association}.

\bibitem[Kennedy, 2023]{kennedy2023towards}
Kennedy, E.~H. (2023).
\newblock Towards optimal doubly robust estimation of heterogeneous causal
  effects.
\newblock {\em Electronic Journal of Statistics}, 17(2):3008--3049.

\bibitem[Kohler and Krzy{\.z}ak, 2005]{kohler2005adaptive}
Kohler, M. and Krzy{\.z}ak, A. (2005).
\newblock Adaptive regression estimation with multilayer feedforward neural
  networks.
\newblock {\em Nonparametric Statistics}, 17(8):891--913.

\bibitem[Kohler and Krzy{\.z}ak, 2016]{kohler2016nonparametric}
Kohler, M. and Krzy{\.z}ak, A. (2016).
\newblock Nonparametric regression based on hierarchical interaction models.
\newblock {\em IEEE Transactions on Information Theory}, 63(3):1620--1630.

\bibitem[Kohler and Langer, 2021]{kohler2021rate}
Kohler, M. and Langer, S. (2021).
\newblock On the rate of convergence of fully connected deep neural network
  regression estimates.
\newblock {\em The Annals of Statistics}, 49(4):2231.

\bibitem[Kreif and DiazOrdaz, 2019]{kreif2019machine}
Kreif, N. and DiazOrdaz, K. (2019).
\newblock Machine learning in policy evaluation: new tools for causal
  inference.
\newblock {\em arXiv preprint arXiv:1903.00402}.

\bibitem[Lenzerini, 2002]{lenzerini2002data}
Lenzerini, M. (2002).
\newblock Data integration: A theoretical perspective.
\newblock In {\em Proceedings of the twenty-first ACM SIGMOD-SIGACT-SIGART
  symposium on Principles of database systems}, pages 233--246.

\bibitem[Li et~al., 2022]{li2022transfer}
Li, S., Cai, T.~T., and Li, H. (2022).
\newblock Transfer learning for high-dimensional linear regression: Prediction,
  estimation and minimax optimality.
\newblock {\em Journal of the Royal Statistical Society Series B: Statistical
  Methodology}, 84(1):149--173.

\bibitem[Li et~al., 2023]{li2023transfer}
Li, S., Cai, T.~T., and Li, H. (2023).
\newblock Transfer learning in large-scale gaussian graphical models with false
  discovery rate control.
\newblock {\em Journal of the American Statistical Association},
  118(543):2171--2183.

\bibitem[Liang and Yu, 2022]{liang2022semiparametric}
Liang, M. and Yu, M. (2022).
\newblock A semiparametric approach to model effect modification.
\newblock {\em Journal of the American Statistical Association},
  117(538):752--764.

\bibitem[Liu et~al., 2021]{liu2021double}
Liu, M., Zhang, Y., and Zhou, D. (2021).
\newblock Double/debiased machine learning for logistic partially linear model.
\newblock {\em The Econometrics Journal}, 24(3):559--588.

\bibitem[Ma and Zhu, 2012]{ma2012semiparametric}
Ma, Y. and Zhu, L. (2012).
\newblock A semiparametric approach to dimension reduction.
\newblock {\em Journal of the American Statistical Association},
  107(497):168--179.

\bibitem[Ma and Zhu, 2013]{ma2013efficient}
Ma, Y. and Zhu, L. (2013).
\newblock Efficient estimation in sufficient dimension reduction.
\newblock {\em Annals of statistics}, 41(1):250.

\bibitem[Maity et~al., 2022]{maity2022minimax}
Maity, S., Sun, Y., and Banerjee, M. (2022).
\newblock Minimax optimal approaches to the label shift problem in
  non-parametric settings.
\newblock {\em Journal of Machine Learning Research}, 23(346):1--45.

\bibitem[Nie and Wager, 2021]{nie2021quasi}
Nie, X. and Wager, S. (2021).
\newblock Quasi-oracle estimation of heterogeneous treatment effects.
\newblock {\em Biometrika}, 108(2):299--319.

\bibitem[Sanchez et~al., 2022]{sanchez2022causal}
Sanchez, P., Voisey, J.~P., Xia, T., Watson, H.~I., O’Neil, A.~Q., and
  Tsaftaris, S.~A. (2022).
\newblock Causal machine learning for healthcare and precision medicine.
\newblock {\em Royal Society Open Science}, 9(8):220638.

\bibitem[Schmidt-Hieber, 2020]{schmidt2020nonparametric}
Schmidt-Hieber, A.~J. (2020).
\newblock Nonparametric regression using deep neural networks with relu
  activation function.
\newblock {\em Annals of statistics}, 48(4):1875--1897.

\bibitem[Sidheekh et~al., 2024]{sidheekh2024credibility}
Sidheekh, S., Tenali, P., Mathur, S., Blasch, E., Kersting, K., and Natarajan,
  S. (2024).
\newblock Credibility-aware multi-modal fusion using probabilistic circuits.
\newblock {\em arXiv preprint arXiv:2403.03281}.

\bibitem[Tian and Feng, 2023]{tian2023transfer}
Tian, Y. and Feng, Y. (2023).
\newblock Transfer learning under high-dimensional generalized linear models.
\newblock {\em Journal of the American Statistical Association},
  118(544):2684--2697.

\bibitem[Tian et~al., 2023]{tian2023learning}
Tian, Y., Gu, Y., and Feng, Y. (2023).
\newblock Learning from similar linear representations: adaptivity, minimaxity,
  and robustness.
\newblock {\em arXiv preprint arXiv:2303.17765}.

\bibitem[Tripuraneni et~al., 2021]{tripuraneni2021provable}
Tripuraneni, N., Jin, C., and Jordan, M. (2021).
\newblock Provable meta-learning of linear representations.
\newblock In {\em International Conference on Machine Learning}, pages
  10434--10443. PMLR.

\bibitem[Wager and Athey, 2018]{wager2018estimation}
Wager, S. and Athey, S. (2018).
\newblock Estimation and inference of heterogeneous treatment effects using
  random forests.
\newblock {\em Journal of the American Statistical Association},
  113(523):1228--1242.

\bibitem[Weiss et~al., 2016]{weiss2016survey}
Weiss, K., Khoshgoftaar, T.~M., and Wang, D. (2016).
\newblock A survey of transfer learning.
\newblock {\em Journal of Big data}, 3:1--40.

\bibitem[Wiens et~al., 2014]{wiens2014study}
Wiens, J., Guttag, J., and Horvitz, E. (2014).
\newblock A study in transfer learning: leveraging data from multiple hospitals
  to enhance hospital-specific predictions.
\newblock {\em Journal of the American Medical Informatics Association},
  21(4):699--706.

\bibitem[Zhang and Yang, 2018]{zhang2018overview}
Zhang, Y. and Yang, Q. (2018).
\newblock An overview of multi-task learning.
\newblock {\em National Science Review}, 5(1):30--43.

\bibitem[Zhou et~al., 2024]{zhou2024doubly}
Zhou, D., Liu, M., Li, M., and Cai, T. (2024).
\newblock Doubly robust augmented model accuracy transfer inference with high
  dimensional features.
\newblock {\em Journal of the American Statistical Association}, pages 1--26.

\end{thebibliography}

\end{document}